\documentclass[letter]{aa} 

\usepackage{graphicx}
\usepackage{comment}
\usepackage{txfonts}
 \usepackage{hyperref}
 \usepackage{multirow}
 \usepackage{xcolor}
 \usepackage{dblfloatfix}

\begin{document}

   \title{Spectropolarimetry of NGC~1275 reveals a narrow-line radio galaxy with polarization parallel to its radio jet axis}

   \titlerunning{A jet-aligned Seyfert-2 galaxy}
   \authorrunning{F. Marin et al.}

   \author{F. Marin\inst{1}          
          \and
          T. Pursimo\inst{2,3}  
          \and
          I. Liodakis\inst{4,5} 
          \and
          E. Lindfors\inst{6} 
          \and
          J. Biedermann\inst{1} 
          \and
          D. Hutsem\'ekers\inst{7}
          \and
          M. Turkki\inst{8,6,9} 
          }

   \institute{Universit\'e de Strasbourg, CNRS, Observatoire Astronomique de Strasbourg, UMR 7550, 11 rue de l'universit\'e, 67000 Strasbourg, France\\
             \email{frederic.marin@astro.unistra.fr}
             \and
             Nordic Optical Telescope, Apartado 474, E-38700 Santa Cruz de La Palma, Santa Cruz de Tenerife, Spain
             \and
             Department of Physics and Astronomy, Aarhus University, Munkegade 120, DK-8000 Aarhus C, Denmark
             \and
             Institute of Astrophysics, Foundation for Research and Technology - Hellas, Voutes, 7110, Heraklion, Greece
             \and   
             Max-Planck-Institut f\"{u}r Radioastronomie, Auf dem H\"{u}gel 69, D-53121 Bonn, Germany
             \and
             Department of Physics and Astronomy, University of Turku, Vesilinnantie 5, Turku FI-20014, Finland
             \and         
             Institut d’Astrophysique et de G\'eophysique, Universit\'e de Li\`ege, All\'ee du 6 Ao\^ut 19c, B5c, 4000 Li\`ege, Belgium
             \and 
             Finnish Centre for Astronomy with ESO (FINCA), University of Turku, FI-20014 Turku, Finland
             \and 
             Aalto University Mets\"{a}hovi Radio Observatory, Mets\"{a}hovintie 114, 02540 Kylm\"{a}l\"{a}, Finland
             }

   \date{Received June 29, 2025; accepted October 7, 2025}

   \abstract
{Concomitant with the Imaging X-ray Polarimetry Explorer (IXPE) observation of the Perseus cluster, we obtained optical spectropolarimetry of its central active galactic nucleus, NGC~1275, using the Alhambra Faint Object Spectrograph and Camera (ALFOSC) on the Nordic Optical Telescope (NOT). While the total-light spectrum confirms its edge-on, core obscured (type-2) classification, the polarized spectrum shows a polarization angle aligned with the arcsecond radio jet axis — an exceptional behavior for type-2 objects. Our polarization analysis also reveals wavelength-dependent linear polarization at level 2-3\% in the continuum, likely rising from a combination of variable synchrotron emission and scattering, supporting a narrow-line radio galaxy classification with characteristics of a jet-dominated source.}


   \maketitle
%

\section{Introduction}
\label{Introduction}

Among the high-density regions of the nearby Universe, the Perseus cluster (situated at about 71~Mpc from us \citealt{Aguerri2020}) is arguably one of the most extensively investigated. Belonging to the Perseus-Pisces supercluster, a large-scale structure in the southern sky extending about 48~Mpc in projected distance \citep{Chincarini1983}, the Perseus cluster contains one of the most remarkable known core-cooling flows, closely associated with the central type-D giant elliptical galaxy NGC~1275 and its active galactic nucleus (AGN, see \citealt{Conselice2001}). At $z$ = 0.017670, NGC~1275's spatial scale is thus 358~pc~arcsec$^{-1}$, with H$_0$ = 70 km s$^{-1}$ Mpc$^{-1}$ \citep{Wilman2005}.

Due to its strong and narrow line emission, filamentary structures, blue color, and nuclear activity, NGC~1275 was first classified as an edge-on, obscured (type-2) Seyfert galaxy \citep{Veron1978}, but it is debatable for many reasons: radio-quiet AGNs are typically spiral galaxies and they are rarely found at the centers of dense galaxy clusters \citep{Best2007}. In addition, its $\sim$ 5-37~GHz radio flux densities (10-60~Jy between 1965 and 2022, \citealt{Paraschos2023}) would rather classify it as a narrow-line radio galaxy, i.e. the radio-loud counterpart of a Seyfert-2 galaxy. Finally, NGC~1275 also shows variability in both optical and radio polarization which would orient its classification towards a "misdirected BL Lacertae type object", i.e., a radio galaxy whose jet is not aligned close to our line of sight \citep{Antonucci1993}. Indeed, NGC~1275 is the source of highly variable, powerful radiation that is relatively bright in radio \citep{Paraschos2023}, X-rays \citep{Churazov2003} and $\gamma$-rays \citep{Abdo2009}. While the radio emission comes from non-thermal mechanisms and the optical counterpart is probably a mix of synchrotron (jet) and thermal (accretion disk, stars) emission, the origin of high-energy radiation remains unsolved.

This is the reason why NGC~1275 was the target of a deep X-ray polarimetric investigation by the Imaging X-ray Polarimetry Explorer (IXPE) in order to assess both the complex structure of the surrounding intracluster medium and the AGN high energy emission mechanisms at once, thanks to the relatively large field-of-view of the detector units (about 12.9 arcmin$^2$) and its angular resolution ($\le$ 28 arcsec, \citealt{Weisskopf2022}). IXPE observed the Perseus cluster for a total of approximately 2.5~Ms between January 26 and March 26, 2025. The results are presented in Liodakis et al. (submitted), at least for the central AGN, observed simultaneously in X-ray, R-band and radio polarimetry. 

In this accompanying paper, we report on an optical spectropolarimetric observation of the same target, acquired during the last days of the IXPE observation. This is the first polarized spectrum ever published for this target and it lead to the discovery of a unique orientation of the polarization position angle with respect to the parsec-scale radio jet of NGC~1275. As these results provide crucial context for the ongoing analysis of the deeper IXPE observations of the Perseus cluster, we consider it important to report them promptly. 

\begin{figure*}[b]
\centering
\includegraphics[width=\textwidth]{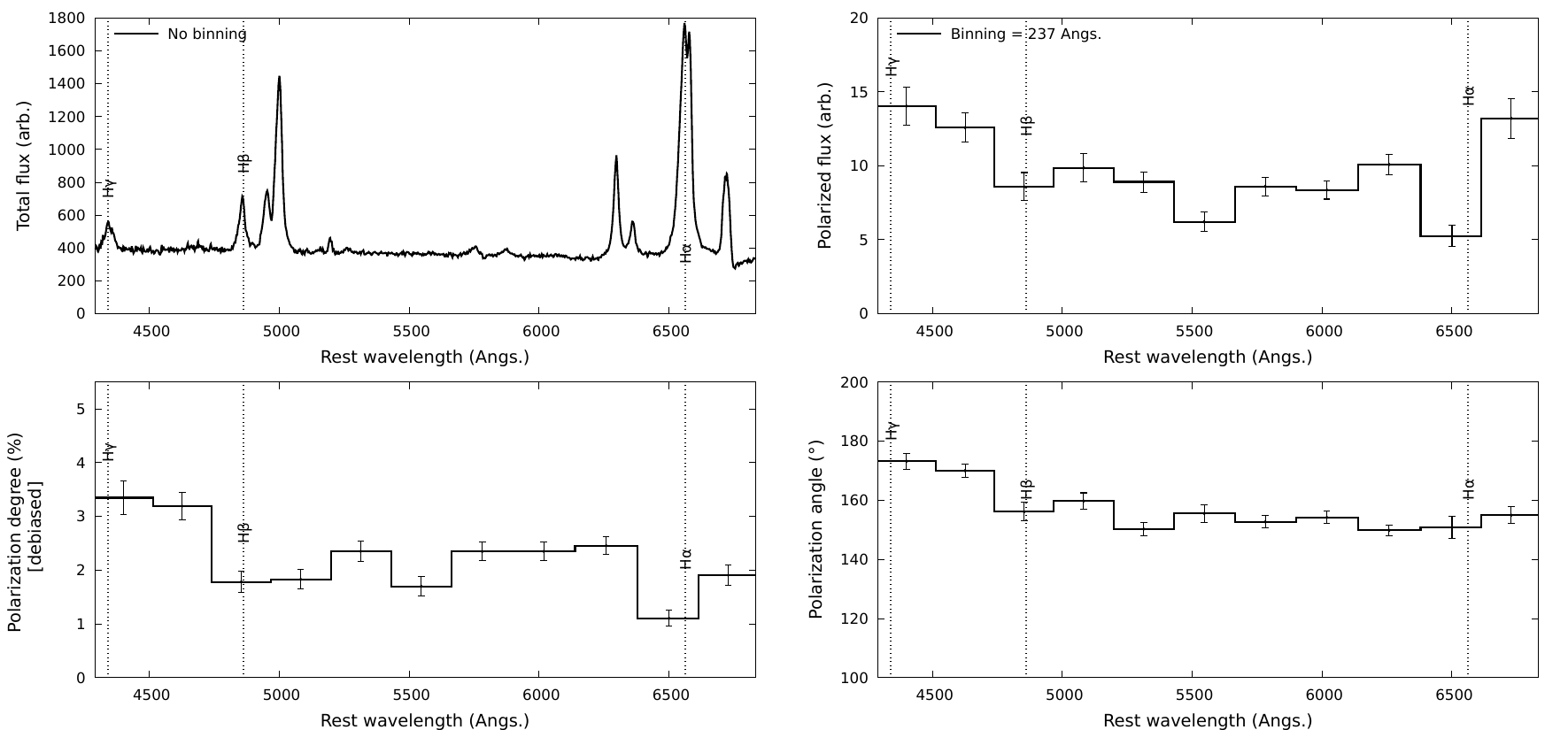}
\caption{NGC~1275 was observed with NOT/ALFOSC on March 21, 2025. The top-left panel shows the total flux spectrum (corrected for instrumental response, in arbitrary units), and the top-right shows the polarized flux (total flux multiplied by the polarization degree). The bottom-left displays the linear polarization degree $P_{\rm d}$, while the bottom-right shows the polarization position angle $\theta$. All but the total flux spectrum were rebinned to 137 consecutive pixels ($\sim$237~\AA), with observational errors indicated per spectral bin.}
\label{Fig:Binned_data}%
\end{figure*}

\section{Observations}
\label{Observation}

Spectropolarimetry of NGC~1275 (3C~84) was obtained on March 21, 2025, using the Alhambra Faint Object Spectrograph and Camera (ALFOSC) mounted on the Nordic Optical Telescope (NOT). Conditions were clear, with 1.5" seeing and airmass $\sim$1.3, minimizing wavelength-dependent flux losses. The ALFOSC spectropolarimetry setup uses a rotating $\lambda$/2 retarder plate, a 1.5"-wide, 15"-long slit aligned with the parallactic angle to minimize slit losses, a calcite block, and grism  \#19. This grism covers 4400–6950~\AA\ with 1.2~\AA/pixel dispersion and a spectral resolution of 970 at 5640~\AA\ (for a 1" slit). The retarder plate was rotated in 22.5$^\circ$ steps over four angles, with 90-second exposures per step (total 360 seconds). Data were reduced with the Python pipeline $PypeIt$ \citep{Prochaska2020}. Due to ALFOSC design, the Wollaston prism’s left and right beams have slightly offset wavelength zero points. One-dimensional spectra at waveplate angles $\alpha$ = 0$^\circ$, 22.5$^\circ$, 45$^\circ$, and 67.5$^\circ$ were corrected for this and rebinned to a common wavelength scale.

From these, we computed the normalized flux differences for each angle and derived the Stokes parameters $Q$ and $U$ using standard formulas, namely $Q = \tfrac{1}{2}(f_{0^\circ} - f_{45^\circ})$ and $U = \tfrac{1}{2}(f_{22.5^\circ} - f_{67.5^\circ})$, where $f_\alpha = -(I_\mathrm{beam1} - I_\mathrm{beam2}) / (I_\mathrm{beam1} + I_\mathrm{beam2})$ for each waveplate angle $\alpha$. The total intensity spectrum $I$ was computed as the sum of all beams. The $Q$ and $U$ spectra were divided by $I$, resulting in the normalized Stokes parameters $u$ and $q$. Errors on $I$, $u$ and $q$ were then computed from the square roots of the spectra (in electron units) and propagated. We computed the linear polarization degree $P$ (= $\sqrt{q^{2}+u^{2}}$) and polarization angle $\theta$ (= $\frac{1}{2}\arctan\,(\frac{u}{q})$) using the standard formulas. $P$ and $\theta$ were finally corrected for the wave plate chromatic dependence following the NOT guidelines.

Unpolarized (HD154892, \citealt{Turnshek1990}) and highly polarized (HD161056, \citealt{Piirola2021}) standard stars were observed shortly after NGC~1275, confirming instrumental polarization below 0.1\%. Their measured values agreed with the tabulated ones within uncertainties \citep{Piirola2020}. Interstellar polarization (ISP) in the direction of the center of the Perseus cluster is low, below 0.2\%, with position-dependent polarization angles \citep{Panopoulou2025}. Because the polarization we are about to report exceeds 1\% in all spectral bins, no interstellar polarization correction was deemed mandatory.

\section{Analysis}
\label{Analysis}

We report in Fig.~\ref{Fig:Binned_data} our NOT/ALFOSC spectropolarimetric observation of NGC~1275. At rest wavelengths, the total flux spectrum extends from 4320 to 6824~\AA. It has a modest signal-to-noise ratio (median SNR across the spectrum $\sim$ 36) and is shown at a spectral resolution of $\sim$ 1.2~\AA. The polarization signal is more noisy and thus the normalized Stokes parameters are only estimates of their true value. To correct this bias, in the following, we refer to the polarization degree as its improved estimator: the debiased polarization degree, given by $P_{\rm d}$ = $\sqrt{P^2-\sigma^2_p}$ \citep{Simmons1985}. In the graphs, $P_{\rm d}$ has been rebinned so that each resolution element has $P_{\rm d}$/$\sigma_P $ $>>$ 7.

\subsection{Spectroscopy}
\label{Analysis:Spectra}

Focusing on the total flux only (see Fig.~\ref{Fig:Binned_data}, top-left), we see that the spectrum of NGC~1275 is characterized by a monotonic continuum (uncorrected from the host galaxy starlight nor reddening) and bright, apparently narrow, nuclear emission lines. Such spectrum is archetypal of type-2 AGNs, where the central engine (the supermassive black hole, its accretion structure and the broad emission line region -- the BLR) is hidden behind an optically thick veil of dust and gas \citep{Antonucci1993}. This spectrum is not surprising, as it was already presented by \citet{Kennicutt1992}. However, a quantitative comparison of the 3650-7100~\AA\, spectrum presented in \citet{Kennicutt1992} with ours is not straightforward as the authors used a 2.5"-wide slit in drift scans along the major axis of the galaxy, with scan lengths 45"-800", thereby integrating a large fraction of the host galaxy and extended filaments. This explains why their spectrum includes spatially-offset narrow-line components, producing multi-peaked [O III] or H$\beta$ profiles if different filaments have different velocities, that do not appear in our observation.

To push further the investigation, we fitted the emission lines with Lorentzian profiles (slightly better suited than Gaussian profiles in the case of AGNs, but the two line profiles are largely interchangeable), after subtracting the continuum. The results of our fits are listed in Tab.~\ref{Tab:Lines}. Focusing first on the H$\alpha$/H$\beta$ ratio -- that is an indicator of the reddening towards NGC~1275 --, we find a value of approximately 3.9. This ratio is significantly higher than the Case B recombination value of $\sim$ 2.86, indicating moderate reddening. Using the Fitzpatrick extinction law \citep{Fitzpatrick1999} and assuming an intrinsic H$\alpha$/H$\beta$ of 2.86, this corresponds to a color excess E(B–V) $\sim$ 0.34. However, since no spectrophotometric calibration was applied and the spectrum could be affected by atmospheric extinction, this estimate as well as those in the following paragraphs should be regarded as indicative rather than precise.

Looking at the widths of the emission lines, we find that almost all of them are smaller than 1000 km~s$^{-1}$, as expected from Seyfert-2 and narrow-line radio galaxies \citep{Netzer1990}. H$\alpha$ and H$\beta$ have consistent full width at half maximums (FWHMs), indicating that they are originating from the same emission region. The fact that the two [O~III] lines have a FWHM superior than 1000 km~s$^{-1}$ is unusual, but this may be due the jet–cloud interactions or intrinsic emission of the cluster's complex network of ionized filaments that encompass the central galaxy \citep{Conselice2001,Rhea2025}. Interestingly, \citet{Goodrich1992} found a similar value for the [O~III] doublet (1291 km~s$^{-1}$). Overall, the heterogeneity of FWHM values suggests a mixture of regions within our integration aperture, some very close to the core, others extended within the narrow line region (NLR).

Considering the integrated line flux ratios [O~III]~$\lambda$5007/H$\beta$ ($\sim$3.3), [N~II]~$\lambda$6583/H$\alpha$ ($\sim$0.5), ([S~II]~$\lambda$6696 + $\lambda$6756)/H$\alpha$ ($\sim$0.2), and [O~I]~$\lambda$6300/H$\alpha$ ($\sim$0.3), we find values consistent with a Seyfert nucleus \citep{Baldwin1981}, showing signs of high excitation and possibly fast shocks (e.g., [O~I]/H$\alpha$ $>$ 0.1), as also suggested by the weak shock fronts propagating outwards the inner part of Perseus detected by \citet{Fabian2003}. However, we note that these line ratios are derived from long-slit spectra, which are subject to several limitations. In particular, the slit orientation, the extraction aperture, and the relatively wide slit likely include emission from physically distinct regions, such that the integrated spectrum may mix gas with different physical conditions. The electron density derived from the [S~II] doublet ratio ($\sim$ 0.7) yields 2 $\times$ 10$^3$ cm$^{-3}$, a fairly typical value for the inner NLR of AGNs \citep{Proxauf2014}. It suggests moderate gas compression, likely from AGN winds or shocks. While this provides a reasonable estimate, it should be interpreted with caution, as gas compression inferred from this measurement may partly reflect averaging over regions with different densities and ionization conditions, possibly influenced by AGN winds or shocks.

\begin{table}
\caption{Emission line (rest-frame air wavelengths) properties}              
\centering                                     
\begin{tabular}{l c c c c}          
\hline\hline                        
Feature & Peak intensity (arb.) & FWHM (km~s$^{-1}$) \\   
\hline                                   
H$\gamma$ (4340~\AA) & (0.093 $\pm$ 0.002)$^a$ & (1980 $\pm$ 38)$^a$ \\
H$\beta$ (4861~\AA) & 0.189 $\pm$ 0.002 & 1314 $\pm$ 21 \\
{[O~III]} (4959~\AA) & 0.214 $\pm$ 0.001 & 1280 $\pm$ 19  \\
{[O~III]} (5007~\AA) & 0.625 $\pm$ 0.004 & 1257 $\pm$ 7 \\
{[N~I]} (5200~\AA) & 0.053 $\pm$ 0.002 & 541 $\pm$ 24 \\
{[N~II]} (5755~\AA) & (0.030 $\pm$ 0.002)$^b$ & (898 $\pm$ 54)$^b$ \\
He~I (5876~\AA) & 0.019 $\pm$ 0.001 & 955 $\pm$ 293 \\
{[O~I]} (6300~\AA) & 0.357 $\pm$ 0.003 & 854 $\pm$ 7 \\
{[O~I]} (6364~\AA) & 0.117 $\pm$ 0.001 & 860 $\pm$ 12 \\
{[N~II]} (6548~\AA) & 0.203 $\pm$ 0.002 & 916 $\pm$ 1 \\
H$\alpha$ (6563~\AA) & 0.736 $\pm$ 0.007 & 1136 $\pm$ 10 \\
{[N~II]} (6583~\AA) & 0.622 $\pm$ 0.008 & 686 $\pm$ 9 \\
{[S~II]} (6716~\AA) & 0.202 $\pm$ 0.004 & 516 $\pm$ 13  \\
{[S~II]} (6731~\AA) & 0.286 $\pm$ 0.004 & 459 $\pm$ 13 \\
\hline                                            
\end{tabular}
\label{Tab:Lines}     
\tablefoot{$^a$ Likely contaminated by [O~III]~4363~\AA; $^b$ Possible blend with [Fe~VII]~5720~\AA. The spectrum was normalized to the peak (strongest line) before continuum subtraction.}
\end{table}

\begin{table}
\caption{Polarization in several bands (rest wavelengths)}              
\centering                                     
\begin{tabular}{l c c}          
\hline\hline                        
Region & $P_{\rm d}$ (\%) & $\theta$ ($^\circ$) \\   
\hline                                   
4320--6824~\AA & 2.17 $\pm$ 0.06 & 157.6 $\pm$ 0.8\\
4400--4780~\AA & 3.23 $\pm$ 0.20 & 164.0 $\pm$ 1.8\\
5086--5675~\AA & 2.02 $\pm$ 0.11 & 154.8 $\pm$ 1.6\\
H$\beta$+{[O~III]} & 1.53 $\pm$ 0.17 & 150.5 $\pm$ 3.0\\
H$\alpha$+{[N~II]} & 1.20 $\pm$ 0.16 & 154.3 $\pm$ 3.8\\
\hline                                            
\end{tabular}
\label{Tab:Pol}     
\end{table}

\subsection{Polarization}
\label{Analysis:Polarization}

Focusing now on the polarized spectra of NGC~1275, we see several interesting features, never reported before. First, looking at the polarized flux spectrum (that is the multiplication of the total flux spectrum with $P$, see Fig.~\ref{Fig:Binned_data}, top-right), we see that the continuum appears wavelength-dependent, with dips of polarized fluxes at the location of the strongest narrow emission lines in comparison to the adjacent bins. It means that these lines are likely less polarized than the adjacent continuum, a common behavior in AGN spectropolarimetry \citep{Antonucci1993}. The low signal-to-noise of our spectrum provides no evidence for the actual presence or absence of broad lines in the polarized flux.

The debiased, linear polarization degree (Fig.~\ref{Fig:Binned_data}, bottom-left) also varies across the spectrum. It is 3.23\% $\pm$ 0.20\% in the blue and 2.02\% $\pm$ 0.11\% in the red bands, see Tab.~\ref{Tab:Pol}. The contribution of the host starlight -- which is essentially unpolarized -- decreases with decreasing wavelength, which could potentially explain why the polarization degree rises shortward of H$\beta$. Another potential explanation is that scattering efficiency drops at longer wavelengths, potentially revealing the emergence of a dominant dust scattering component. Anyhow, both bands show significant polarization (well above errors), indicating that this is a real effect rather than noise.

The polarization angle $\theta$ (Fig.~\ref{Fig:Binned_data}, bottom-right) shows a similar trend, averaging 164.0$^\circ$ $\pm$ 1.8$^\circ$ in the blue and 154.8$^\circ$ $\pm$ 1.6$^\circ$ in the red band (see Tab.~\ref{Tab:Pol}). It thus rotates by 9.2$^\circ$ $\pm$ 2.4$^\circ$ between 4400–4780~\AA\, and 5086–5675~\AA, a wavelength dependence significant at the 3.8$\sigma$ level that can reflect the relative contributions or intrinsic properties of different emitters and/or scatterers.

\citet{Antonucci1984} reported a spectropolarimetric measurement of the same target, which would make it the first spectropolarimetric observation of NGC~1275, but no spectra were shown. Only an integrated polarization of $P$ = 0.85\% $\pm$ 0.10\% at $\theta$ = 120$^\circ$ $\pm$ 3$^\circ$ is quoted, but the author judged it to be spurious, as strongly affected by ISP. We now know that ISP is less than 0.2\% and the low $P$ recorded at the time was likely due to a combination of intrinsic variability and methodological differences (e.g. larger aperture increasing the host galaxy contribution).

The [O~III] doublet polarization we measure (see Tab.~\ref{Tab:Pol}) agrees within uncertainties with the value reported by \citet{Goodrich1992}: 1.87\% $\pm$ 0.39\% at 149$^\circ$ $\pm$ 6$^\circ$ in the forbidden lines. This suggests the [O~III] polarization has remained stable over 35 years. According to the author, the lines are essentially polarized by transmission through aligned dust grains \citep{Antonucci1985}, which would explain their temporal stability.

\section{Discussion and conclusions}
\label{Conclusions}

We have found that the optical polarized spectrum of NGC~1275 displays typical type-2 AGN features: narrow emission lines only, strong [O~III], [N~II] and [S~II] features indicative of AGN photoionization and/or shock excitation (which are common in cooling flow filaments), moderate polarization degree in the continuum and lower polarization in the lines, all common signatures of Seyfert-2 and radio-loud narrow line galaxies, in which the core thermal emission is scattered onto the polar outflows towards the observer. However, two elements don't fit the frame: first, the optical polarization degree of NGC~1275 is known to be variable on weekly time scales \citep{Babadzhanyants1972,Shkodkina2025}; second, there is the issue of the alignment between the continuum polarization angle and the radio jet axis. 

Focusing on the first point, archival data show that the R-band polarization from NGC~1275 varies from 1 to 6\% (Liodakis et al. submitted), with extreme $\theta$ rotations, yet with a time-averaged value of 151$^\circ$ (34$^\circ$ standard deviation). Such variation are typical of BL Lacertae-like objects \citep{Martin1976,Hovatta2016}, i.e. blazing radio-loud AGNs where the observed optical polarization comes from synchrotron emission. Polarization variability is the tell-tale signature of relativistic electrons gyrating in a magnetic field. However, NGC~1275’s modest polarization levels and the presence of line polarization across forbidden and Balmer lines suggest synchrotron emission isn't the sole contributor. While depolarization of a polarized continuum by strong unpolarized lines can indeed mimic line polarization -- as is often seen in blazars \citep{Schutte2022} --, the fairly consistent polarization angle ($\sim$150$^\circ$) across both continuum and lines points toward an additional scattering component with a stable geometry.

Focusing on the second issue brings a possible solution to light. 73~cm (0.4~GHz) radio continuum observations of NGC~1275 revealed a radio jet extending at a position angle of $\sim$160$^\circ$ on arcsecond scales \citep{Pedlar1983}. However, subsequent high-resolution radio observations have shown that the jet orientation is strongly scale-dependent: $\sim$171$^\circ$ at 10 mas ($\sim$ 4 pc) \citep{Venturi1993}, about -147$^\circ$ at a few mas \citep{Plavin2022}, and even sub-mas bending with RadioAstron \citep{Savolainen2023}. This variability makes the direct comparison between the optical polarization angle reported here ($\sim$158$^\circ$) and a fixed radio jet position angle complex. Nonetheless, the apparent alignment between the two on similar, arcsecond scales suggests that synchrotron emission from the jet strongly contributes to the observed optical polarization, probably superimposed on a background of nuclear photons scattered by electrons or dust in the NLR (as revealed by the change in $P$ and $\theta$ in the blue band. In this tentative scenario, the combination of synchrotron and scattering origins could naturally account for both the temporal variability observed in broadband polarimetry and the spectral features identified in our data.

However, a critical question that remains unanswered is whether NGC~1275 possesses a BLR \citep{Punsly2018}. Answering this question would help determine whether the observed radio-to-X-ray continuum emission is entirely due to jet-related synchrotron processes, or if there are additional contributions (e.g., from the accretion disk, corona, or thermal gas; see \citealt{Marin2025} for an example). This requires longer, deeper polarimetric observation in the same band in order to reach sufficient SNR to detect the presence or absence of broad lines in polarized flux and properly measure the polarization in the narrow components. In addition, time-resolved spectropolarimetry could help separate variability from structural effects and quantify the scattered-to-synchrotron ratio.

\begin{acknowledgements}
We would like to deeply thank the anonymous referee for his/her constructive comments. Based on observations made with the Nordic Optical Telescope, owned in collaboration by the University of Turku and Aarhus University, and operated jointly by Aarhus University, the University of Turku and the University of Oslo, representing Denmark, Finland and Norway, the University of Iceland and Stockholm University at the Observatorio del Roque de los Muchachos, La Palma, Spain, of the Instituto de Astrofisica de Canarias. The NOT data were obtained under program ID 70-414. The data presented here were obtained with ALFOSC, which is provided by the Instituto de Astrofisica de Andalucia (IAA) under a joint agreement with the University of Copenhagen and NOT. I.L was funded by the European Union ERC-2022-STG - BOOTES - 101076343. Views and opinions expressed are however those of the authors only and do not necessarily reflect those of the European Union or the European Research Council Executive Agency. Neither the European Union nor the granting authority can be held responsible for them. DH is research director at the F.R.S-FNRS, Belgium. M. Turkki was supported by Research Council of Finland project 362571.
\end{acknowledgements}

\bibliographystyle{aa} 
\bibliography{Bibliography} 

\end{document}